\documentclass{elsart}

\usepackage{epsfig}

\begin{document}
\begin{frontmatter}

\title{Melting and Equilibrium Shape of Icosahedral Gold Nanoparticles}

\author[roch]{Yanting Wang},
\author[roch]{S. Teitel\corauthref{cor}},
\corauth[cor]{Corresponding author: Department of Physics and Astronomy, University of
Rochester, Rochester, NY 14627; Fax: (585) 273-3237}
\ead{stte@pas.rochester.edu}
\author[vienna]{Christoph Dellago}
\address[roch]{Department of Physics and Astronomy, University of
Rochester, Rochester, NY 14627}
\address[vienna]{Institute for Experimental Physics, University of
Vienna, Boltzmanngasse 5, 1090 Vienna, Austria}

\begin{abstract}
We use molecular dynamics simulations to study the melting of gold 
icosahedral clusters of a few thousand atoms.  We pay particular
attention to the behavior of surface atoms, and to the equilibrium 
shape of the cluster.  We find that the surface of the cluster does not 
pre-melt, but rather 
remains ordered up to the melting $T_{\rm m}$.
However the increasing mobility of
vertex and edge atoms significantly soften the surface structure, leading 
to inter- and intra-layer diffusion, and shrinking of the average 
facet size, so that the average shape of the cluster is nearly spherical 
at melting.
\end{abstract}

\begin{keyword}
Clusters\sep gold nanocrystals\sep   molecular dynamics
\PACS 05.45.-a\sep  79.60.-I

\end{keyword}

\end{frontmatter}


Gold particles consisting of tens to thousands of atoms have unique
optical and mechanical properties and hold great promise as building
blocks for nano-bioelectronic devices \cite{HUTCHISON,WILLNER},
catalysts \cite{BELL}, and sensors \cite{MURPHY}. It is therefore
natural that the physics and chemistry of these materials are a
current research subject of great interest \cite{FELDHEIM}. For
future applications knowledge of the structure and stability of gold
nanoparticles of different size and morphology is particularly
important. 

While bulk gold has an fcc crystal structure, the competition between
bulk and surface energies in nanometer sized gold crystallites can result 
in several different competing structures \cite{IIJIMA,LANDMAN}. Depending
on cluster size and external conditions transitions between these
structures have been observed \cite{BARRAT,LANDMAN}.  One such structure which 
has been observed both in simulations \cite{BARTELL,NAM} 
and in experiments \cite{MARKS_REVIEW,ASCENCIO}, is
the ``Mackay icosahedron''\cite{MACKAY,SHELLS}, 
consisting of 20 slightly distorted
fcc tetrahedra, with four $\{111\}$ faces each, meeting at the 
center to form an icosahedral shaped cluster.  The
internal faces of the tetrahedra meet at strain inducing
twin grain boundaries with hcp structure, 
leaving the cluster with 20 external $\{111\}$
facets.  Theoretical models \cite{WULFF,HERRING,INO,CLEVELAND_ZPD} have 
predicted different limits for the stability of such icosahedral 
clusters, and it is unclear whether their formation is an equilibrium 
or rather a kinetic process 
\cite{MARKS_REVIEW,INO,CLEVELAND_ZPD,CLEVELAND_PRL,GARZON,MARKS_PHIL}.  
Nevertheless, it is natural to suppose that 
formation of this structure is related to the very high stability of 
the $\{111\}$ external surfaces.  
Simulations \cite{CARNEVALI} and experiments \cite{GROSSER} on bulk slab-like 
geometries with exposed $\{111\}$ surfaces have shown that, unlike the 
$\{100\}$ and $\{110\}$ surfaces which melt below the bulk melting 
temperature $T_{\rm m}$, the $\{111\}$ surface neither melts nor roughens 
but remains ordered up to 
and above $T_{\rm m}$, and can in fact lead to superheating of 
the solid \cite{DITOLLA}.  
In light of this observation it is interesting to consider how
the high stability of the $\{111\}$ facets effects the melting and
equilibrium shape of such icosahedral nanoclusters.

In order to address this issue, we have performed detailed 
numerical simulations of icosahedral gold nanoclusters of a few thousand atoms, 
obtained by cooling from the melt.  We
pay particular attention to the behavior of the surface atoms and 
to the equilibrium shape.  We find a sharp first order melting transition
$T_{\rm m}$. Unlike earlier results on smaller cuboctahedral clusters
\cite{EAT}, which include non $\{111\}$ facets that pre-melt below
$T_{\rm m}$, we find no surface pre-melting of the  $\{111\}$
facets of our icosahedral cluster.
Nevertheless, we find that there is a considerable 
softening of the cluster surface roughly $\sim 200$ K below $T_{\rm m}$ due 
to the motion of atoms along the vertices and edges of the 
cluster.  In this region we find both intra-layer and 
inter-layer diffusion of atoms, which increases considerably as $T_{\rm m}$ is 
approached.  The equilibrium shape progresses from fully faceted, to 
faceted with rounded edges, to nearly spherical just below $T_{\rm m}$.
Throughout this region, the interior atoms of the cluster remain 
essentially perfectly ordered, until $T_{\rm m}$ is reached.  

Using the many-body ``glue'' potential \cite{ERCOLESSI}
to model interactions among gold atoms, we carry out molecular dynamics 
simulations, integrating the classical equations of motion with the 
velocity Verlet algorithm \cite{SMIT} with a time step of $4.3$ fs.
The results presented here are for a $2624$ atom cluster, with a diameter of $\sim 20$  \AA, but we have 
also considered other sizes.  We start our simulations at a high 
$T=1500$ K $>T_{\rm m}$, and cool using the Andersen thermostat method \cite{SMIT}
to $1000$ K.  We then cool, in intervals of $100$ K, down to $200$ K,
using $5\times 10^{6}$ steps ($21$ ns) at each temperature.  Even though
our $N=2624$ atoms is {\it not} a ``magic number" for a perfect icosahedral
structure (the nearest such number being $2868$), the 
cluster structure we find with this method is nevertheless clearly a  Mackay 
icosahedron consisting of slightly distorted tetrahedra with different number of atoms,
but with a missing central atom. Similar icosahedral structures were obtained in runs
with different particle numbers. Such a central vacancy, postulated for 
copper and aluminum but not for gold \cite{LEGRAND}, has been considered 
previously \cite{BOYER} as a means of partially relieving
the strain caused by the hcp twin grain boundaries between the fcc tetrahedra.

\begin{figure}
\epsfxsize=12truecm
\epsfbox{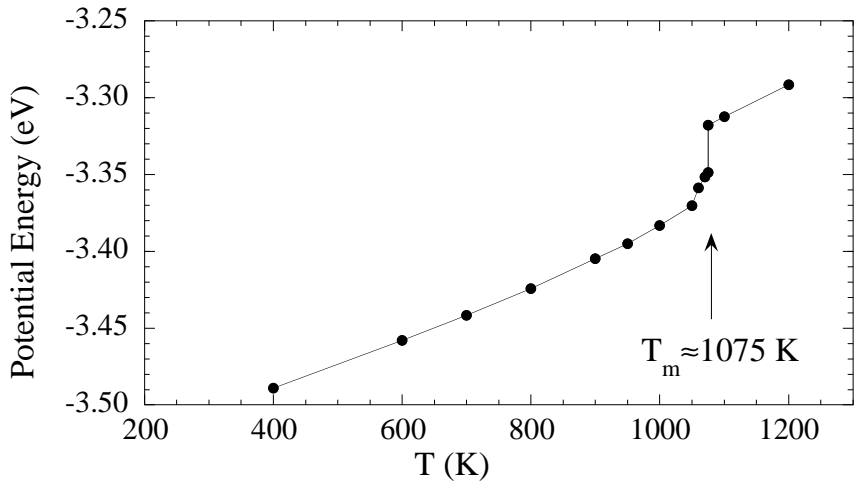}
\caption{Potential energy vs. $T$ for a $2624$ atom gold cluster.  
The sharp jump at $T=1075$ K indicates the first order melting 
transition.
}
\label{f1}
\end{figure}

To study melting and the equilibrium shape, we next heat the cluster up using 
constant temperature molecular dynamics\cite{WANG_DELLAGO}.  
To compute equilibrium 
properties, we take at each temperature $10^{6}$ steps ($4.3$ ns) for
equilibration, followed by  $10^{7}$ steps ($43$ ns) to compute averages. 
We take fine temperature increments in the 
vicinity of the cluster melting transition.  In Fig.\,\ref{f1} we show 
our results for the average potential energy vs. temperature.  We 
found the cluster to melt at $T_{\rm m}=1075$ K, with a discontinuous jump in 
potential energy. Over the whole temperature range from 200 K to 1200 K
gold atoms were never observed to evaporate from the cluster.   

\begin{figure}
\epsfxsize=12truecm
\epsfbox{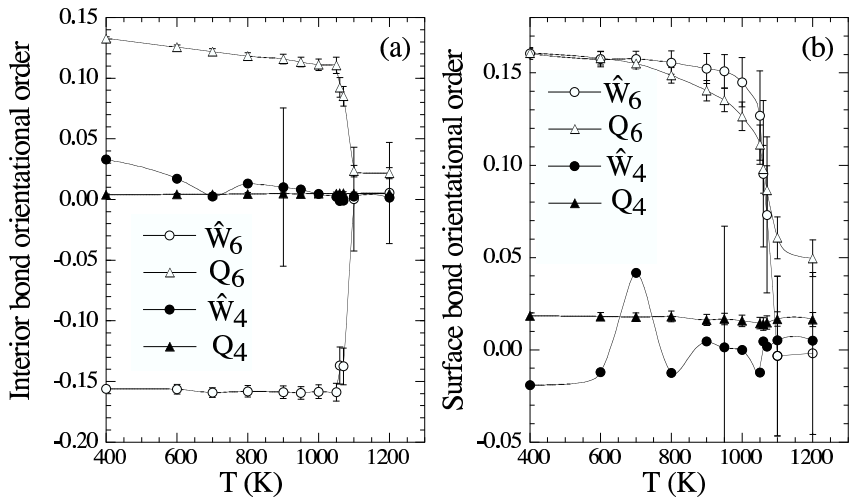}
\caption{Bond orientational order parameters, $Q_{6},\hat 
W_{6},Q_{4},\hat W_{4}$ vs. $T$, averaged over (a) bonds between 
interior atoms only, and (b) bonds between surface atoms only.  For 
$\hat W_{4}$ only a few representative error bars are shown for the 
sake of clarity.
}
\label{f2}
\end{figure}

To characterize the structure of the cluster, we 
measure the standard bond orientational order parameters 
$Q_{6}$, $\hat{W}_{6}$, $Q_{4}$ and $\hat W_{4}$ \cite{STEINHARDT}
which are often used to 
distinguish between different phases of condensed materials \cite{WANG_DELLAGO,WOLDE}. 
These bond order parameters, designed to probe the degree and type of 
crystallinity, are sensitive to the orientational correlations of 
``bonds", i.e. the vectors joning pairs of neighboring atoms.  In the liquid 
phase, such correlations decay quickly with growing distance and the 
bond order parameters vanish. In crystalline solids, on the other hand, 
orientational bond correlations persist over large distances leading to 
order parameters with finite values. 
We refer the reader to the work of Ref.~\cite{STEINHARDT} for the
definition of these quantities, and their values in common crystal structures.  
To distinguish between 
bulk and surface behavior, we first identify all atoms on the surface of 
the cluster, then the atoms in the first sub layer below the surface, and 
so on; the cluster has $9$ such layers. 
We label as ``interior'' atoms those lying below the fourth sub layer
(our results are essentially unchanged if we define the ``interior" as all
atoms below the first sub layer).
In Fig.\,\ref{f2}$a$ we plot the bond order parameters averaged over 
only bonds between the interior atoms.  
We see $Q_{4}, \hat W_{4}\approx 0$ at all $T$, 
while $Q_{6}$ and $\hat W_{6}$ take finite values for $T<T_{\rm m}$
appropriate to the Mackay 
icosahedron.  $Q_{6}$ and $\hat W_{6}$ remain essentially constant, 
decreasing only slightly just below $T_{\rm m}$, indicating that the bulk 
ordering remain stable up until melting. In contrast, 
Fig.\,\ref{f2}$b$ shows the bond order parameters averaged over only 
bonds between the surface atoms.  Again $Q_{4}, \hat W_{4}\approx 0$ at all 
$T$, while $Q_{6}$ and $\hat W_{6}$ take finite values for $T<T_{\rm m}$. 
However here we see a much more pronounced decrease, particularly in $Q_{6}$ 
starting well below $T_{\rm m}$, until both vanish at melting (the finite value 
of $Q_{6}$ above melting is a finite size effect that decreases as 
the cluster size increases).  Thus, while the surface remains ordered below 
$T_{\rm m}$, 
i.e. $|Q_{6}|,|\hat W_{6}|>0$,  the surface order softens to 
a much greater extent than does the bulk as $T_{\rm m}$ is approached.

\begin{figure}
\epsfxsize=12truecm
\epsfbox{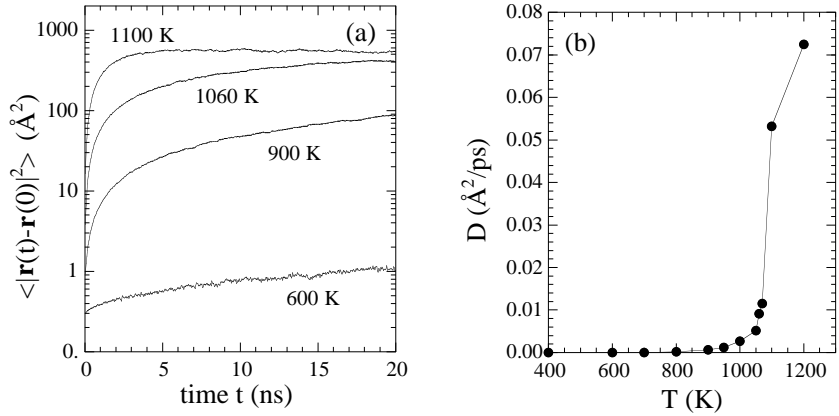}
\caption{(a) Mean square displacement for atoms initially on the 
surface, at various temperatures; (b) Diffusion constant $D$ vs. $T$
for atoms initially on the surface.
}
\label{f4}
\end{figure}

Next we consider the diffusion of the surface atoms.  In 
Fig.\,\ref{f4}$a$ we plot, for several different temperatures,
the average mean square displacement
$\langle |\Delta{\bf r}(t)|^{2}\rangle\equiv
\langle |{\bf r}(t)-{\bf r}(0)|^{2}\rangle$ vs. time $t$, where the 
average is over all the atoms which were initially on the surface of 
the cluster.  
At $T=600$ K, diffusion is very slow, with displacements
after $20$ ns remaining less than one atomic separation.
At $T=900$ K, almost $200$ K below $T_{\rm m}$, diffusion is 
significant.  At $T=1060$ K, $15$ K below $T_{\rm m}$, the mean square 
displacement saturates at large $t$, indicating that atoms now diffuse 
the entire length of the cluster.  Fitting the early time linear part 
of these curves, $\langle |\Delta{\bf r}(t)|^{2}\rangle\sim 6Dt$, we 
plot the diffusion constant in Fig.\,\ref{f4}$b$.

\begin{figure*}[t]
\epsfxsize=5.5truein
\epsfbox{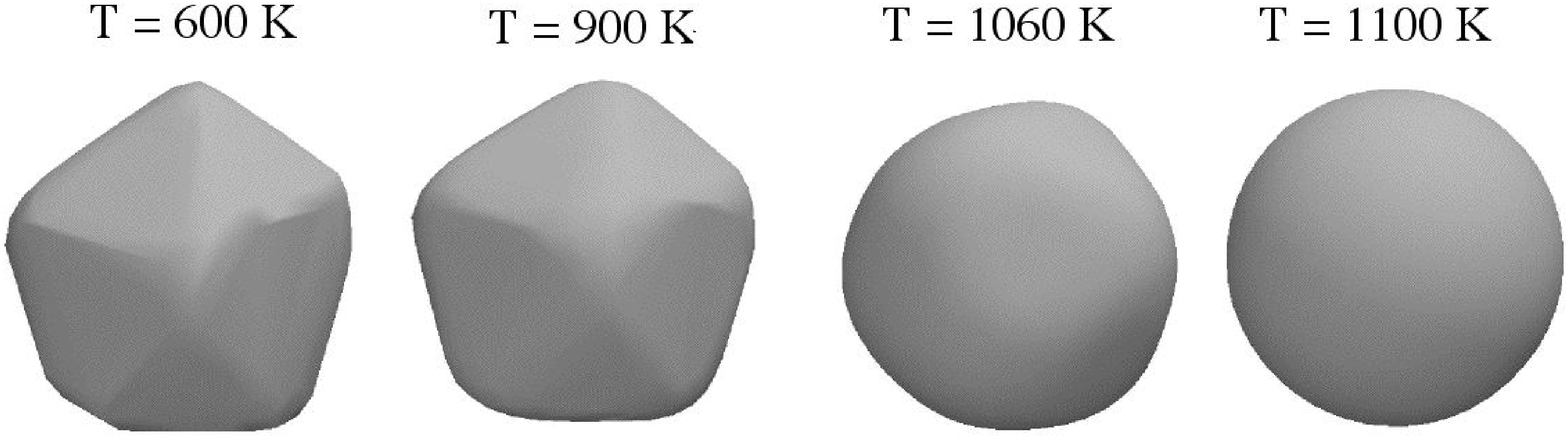}
\end{figure*}
\begin{figure*}[!]
\epsfxsize=5.5truein
\epsfbox{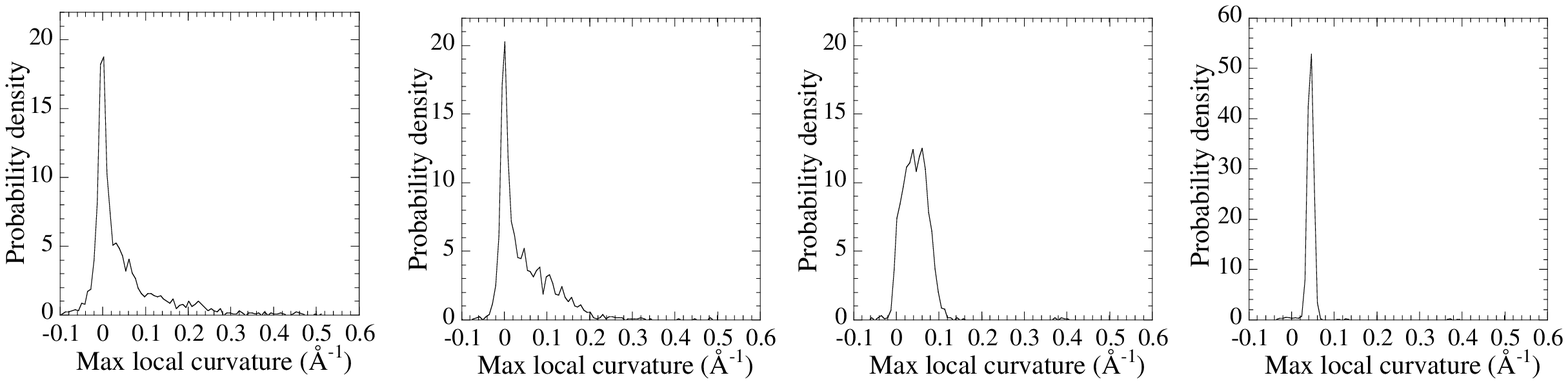}
\end{figure*}
\begin{figure*}[!]
\epsfxsize=5.5truein
\epsfbox{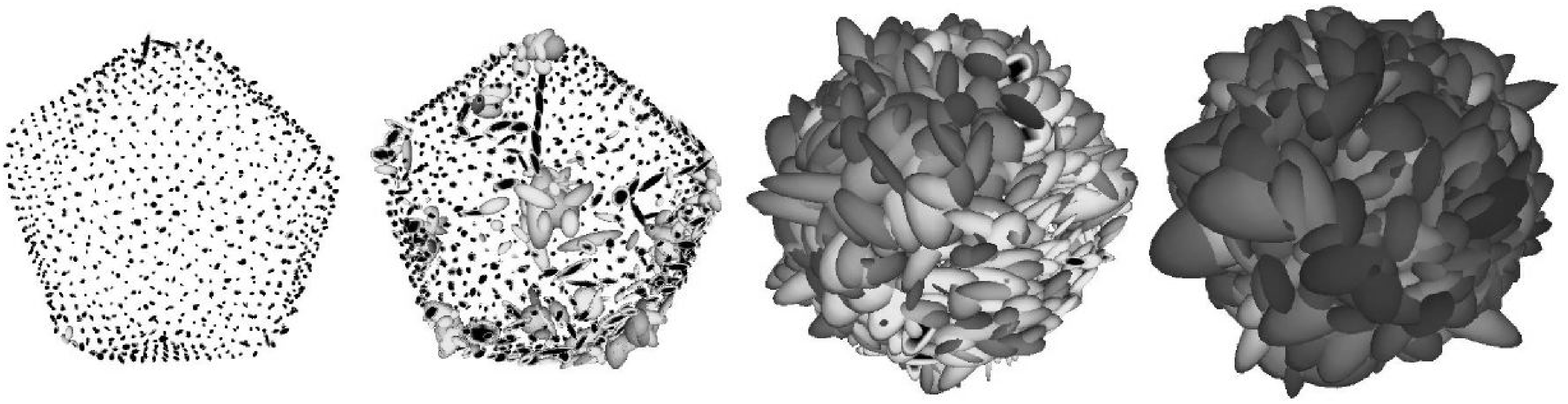}
\caption{For the indicated temperatures, top row: average cluster 
shape for a simulation time of $43$ ns; 
middle row: histogram of the maximal local curvatures of the 
average shape; bottom row: ellipsoids indicate root mean square 
displacements of atoms on the cluster surface over a simulation time 
of $1.075$ ns.
}
\label{f5}
\end{figure*}

The question thus 
arises how to reconcile this observed surface diffusion below $T_{\rm m}$
with the absence of surface melting that is indicated by the finite bond 
orientation order parameters of Fig.\,\ref{f2}$b$.  One possibility is that, as the 
temperature approaches $T_{\rm m}$, all surface atoms become more mobile
but translational order is maintained due to the presence of a periodic 
substrate formed by the ordered sub-layers below the surface. 
Our simulations, however, point to a different picture 
(see last row of figures shown in Fig.\,\ref{f5}).  
For each atom initially on the surface of the cluster, we compute its average 
position $\bar{\bf r}\equiv \langle{\bf r}\rangle$, and its 
average displacement 
correlation matrix $d_{\alpha\beta}\equiv\langle(r-\bar r)_{\alpha}(r-\bar r)_{\beta}\rangle$, where 
$\alpha,\beta=x,y,z$ and the $\langle\ldots\rangle$ stand for 
averages over $25$ configurations, sampled every $43$ ps
of simulation time. 
Taking the eigenvectors of $d_{\alpha\beta}$
and the square root of their corresponding eigenvalues to define
the axes and principal radii of an ellipsoid, gives a convenient 
representation for the root mean square displacement of the atom.
In the last row of Fig.\,\ref{f5} we plot these ellipsoids for each 
atom initially on the surface, centering the ellipsoid at the average 
position of the atom $\bar{\bf r}$.  We show such plots for the same 
temperatures as in Fig.\,\ref{f4}$a$.  We clearly see that 
for $T=600$ K and $900$ K, the 
biggest ellipsoids are at the vertices and edges, while those
for atoms in the facets are in general smaller.  If we let the time 
that we average over increase, we find that the ellipsoids at the vertices and 
edges grow in size, corresponding to diffusion, while those in the 
middle of the facets stay approximately the same, corresponding to 
thermal vibrations without diffusion.  
A more detailed analysis shows that significant interlayer exchange takes 
place between the two topmost layers as much as $200$ K below $T_{\rm m}$. 
In Fig.\,\ref{f5} one can see ellipsoids oriented 
normal to the cluster surface, corresponding to this interlayer diffusion.
For the higher temperatures, $T=1060$ K just below melting, and 
$T=1100$ K just above melting, diffusion is pronounced throughout the 
entire surface.

Finally, we consider the effects of the vertex and edge atom 
diffusion on the equilibrium shape of the cluster.  Because of the small
cluster size, the instantaneous shapes fluctuate significantly at high
temperatures.  But since our simulation algorithm conserves angular 
momentum, and the angular momentum is zero, our sample does not 
rotate as a whole.  Hence we can compute a well defined average equilibrium shape
at each temperature. 
We measure this equilibrium shape by averaging over the instantaneous
shapes as follows.  We divide space up into $842$ approximately equal 
solid angles \cite{HARDIN}, corresponding roughly to the number of surface 
atoms.  We then average the position of the surface atoms found in each 
solid angle over $1000$ configurations sampled at equal times throughout 
the simulation of total time $43$ ns.  This defines the average 
radial position of the cluster within each solid angle, and hence the 
average cluster shape.  In the top row of Fig.\,\ref{f5} we show the 
resulting equilibrium shapes for several temperatures.  We see that 
the shape is rounding out as the temperature increases, assuming a 
nearly perfect spherical shape above $T_{\rm m}$.  

To quantify this, we compute the curvature distribution of the surface 
as follows.  Using the average position of 
the surface in a given solid angle and its nearest neighbors, we fit 
to determine the best tangent plane to these points.  Defining the 
normal to this plane as the $z$ axis, we then find the best paraboloid 
that fits through the points.  The principal curvatures of this 
paraboloid then give our approximation for the two principal 
curvatures of the surface at the given solid angle.  
We define $\kappa$ to be the maximum of these two principal 
curvatures.  In the middle row of Fig.\,\ref{f5} we plot
histograms of $\kappa$ as one varies over all the solid angles
defining the average surface.  At $T=600$ K and 
$900$ K we see a sharp peak at $\kappa=0$ corresponding to the points 
on flat facets, and a high $\kappa$ tail
corresponding to higher curvatures on the edges and vertices.  
At $T=1060$ K, just below $T_{\rm m}=1075$ K, the peak at $\kappa=0$ has
essentially vanished, and one has a broad distribution centered about
$\kappa\simeq 1/R$ with $R=21.5$ \AA,  the radius of the spherical liquid 
drop above $T_{\rm m}$.  At $T=1100$ K, just above 
$T_{\rm m}$, the distribution becomes very sharply peaked about 
$\kappa=1/R$.

To conclude, we have found that gold nanoparticles of a few thousand 
atoms form a Mackay icosahedral structure, with missing central atom, 
when cooled from a liquid.  Upon slow heating, we find that this bulk
structure remains stable up to a sharp first order melting.  The surface 
remains ordered with no pre-melting below $T_{\rm m}$, 
however it softens considerably with increasing diffusion due to mobile
vertex and edge atoms.  As $T_{\rm m}$ is approached, this diffusion 
of edge atoms leads to significant shrinkage of the $\{111\}$ facet
sizes in the average cluster shape, leading to an almost spherical 
shape just below $T_{\rm m}$.  In addition to the cluster of $2624$ 
atoms reported upon here, we have also considered clusters of 
different sizes with $603$ and $1409 $ particles.  While the melting 
temperature $T_{\rm m}$ was observed to increase with increasing 
cluster size, we continued to find the same general features, with 
the surface softening tracking the increase in $T_{\rm m}$.  
It would be interesting to 
know how this surface softening is related to morphological transitions 
observed in gold nanorods at temperatures below the melting temperature 
\cite{WANG_DELLAGO,ELSAYED}. 

This work was funded in part by DOE grant DE-FG02-89ER14017.


%


\end{document}